\documentstyle[12pt]{article}
\begin{document}

\thispagestyle{empty}
\begin{flushright}
FIAN/TD/15-05
\end{flushright}

\vspace{1cm}

\begin{center}

{\bf \Large Secondary Fields in $D>2$ Conformal Theories}

\vspace{1.2cm}

V.N.Zaikin$^*$ and M.Ya.Palchik$^\dagger$

\vspace{6mm}

$^*${\it I.E.Tamm Theoretical Department, P.N.Lebedev Physical Institute, \\
Leninsky prospect 53, 119991, Moscow, Russia}

\vspace{3mm}

$^\dagger${\it Institute of Automation and Electrometry,\\
prospect akademika Koptyuga 1, 630090, Novsibirsk, Russia}

\vspace{1cm}

\begin{abstract}
We consider the secondary fields in $D$-dimensional space,
$D\ge3$, generated by the non-abelian current and energy-momentum
tensor. These fields appear in the operator product expansions
$j^{a}_\mu(x)\varphi(0)$ and $T_{\mu\nu}(x)\varphi(0)$. The
secondary fields underlie the construction proposed herein
(see~[1,2] for more details) and aimed at the derivation of exact
solutions of conformal models in $D\ge3$. In the case of $D=2$
this construction leads to the known~[5] exactly solvable models
based on the infinite-dimensional conformal symmetry.  It is shown
that for $D\ge3$ the existence of the secondary fields is governed
by the existence of anomalous operator contributions (the scalar
fields $R_j$ and $R_T$ of dimensions $d_j = d_T = D-2$) into the
operator product expansions $j^{a}_\mu j^{b}_\nu$ and $T_{\mu\nu}
T_{\rho\sigma}$. The coupling constant between the field $R_j$ and
the fundamental field is found. The fields $R_j$ and $R_T$ are
shown to beget two infinite sets of secondary tensor fields of
canonical dimensions $D-2+s$, where $s$ is the tensor rank.  The
current and the energy-momentum tensor belong to those families,
their Green functions being expressed through the Green functions
of the fields $R_j$ and $R_T$ correspondingly. We demonstrate that
the Ward identities give rise to the closed set of equations for
the Green functions of the fields~$R_j$ and~$R_T$.
\end{abstract}

\end{center}

\date{}
\let\oldsect=\section
\def\section{\setcounter{equation}{0}\oldsect}
\def\theequation{\thesection.\arabic{equation}}
\def\rinv{\mathrel{\mathop{\rightarrow}\limits^R{}}}
\def\res{\mathop{\rm res}}

\newpage
\setcounter{page}{1}

\section{Introduction and Discussion of the Results}
The articles~[1,2] (and the references therein) discuss the
construction enabling one to derive a family of exactly solvable
models in conformal quantum field theory in $D$-dimensional space,
$D\ge2$. This construction is based upon the following feature of
conformal theory: each fundamental complex field $\varphi(x)$ may
be associated (provided that the certain assumptions are made, see
below) with two infinite sets  of secondary tensor fields~[1,2]:
\begin{equation}
P_s(x),P_s^{(s_1)}(x),P_s^{(s_1s_2)}(x),\ldots,P_s^{(s_1\ldots
s_k)}(x),\ldots
\end{equation}
of dimensions
$$l_s = d+s, \quad s = 1,2,\ldots$$

Here $d$ is the dimension of the fundamental field $\varphi(x)$,
$s$ is the tensor rank.  One of these sets of fields is associated
with the conserved current $j_\mu(x)$, while the other is
associated with the energy-momentum tensor $T_{\mu\nu}(x)$.  The
secondary fields~(1.1) appear in  the operator product expansions:
\begin{eqnarray}
&&\kern-40pt j_\mu(x)\varphi(0) = \sum_s [P_s^j],\ j_\mu(x)
P_{s_1}^j(0) = \sum_s [P_s^{j(s_1)}],\
j_\mu P_{s_2}^{j(s_1)}(0) = \sum_s [P_s^{j(s_1s_2)}]\ldots, \\
&&\kern-40pt T_{\mu\nu}(x)\varphi(0) = \sum_s [P_s^T],\
T_{\mu\nu}(x) P_{s}^T(0) = \sum_s [P_s^{T(s_1)}],\ T_{\mu\nu}(x)
P_{s_2}^{T(s_1)}(0) = \sum_s [P_s^{T(s_1s_2)}]\ldots
\end{eqnarray}
and may be found from the Ward identities for the Green functions
$\langle j_\mu\varphi\ldots\rangle$ and $\langle
T_{\mu\nu}\varphi\ldots\rangle$

 Here we use the standard notation
$[P]$ (see~[1] for example) for the total contribution of the
field $P(x)$ and all its derivatives.  What is essential that in
any $D$-dimensional conformal theory  the Green functions $\langle
j_\mu\varphi\ldots\rangle$ and $\langle
T_{\mu\nu}\varphi\ldots\rangle$ are uniquely determined~[1,2] from
the Ward identities provided that the fields $j_\mu$ and
$T_{\mu\nu}$ are conformally irreducible\footnote{The conventional
conformal transformations ($j_\mu(x)\rinv
(x^2)^{-D+1}g_{\mu\nu}(x)j_\nu(Rx)$ where $Rx=x_\mu/x^2$ and
analogously for the energy-momentum tensor $T_{\mu\nu}$) in $D>2$
give rise to irreducible representations of the conformal group,
see~[3,4]. The requirements for $j_\mu,T_{\mu\nu}$ and their Green
functions which single out the irreducible representations are
discussed in~[1,2] and, in more detail, in~[6].}. In addition to
irreducibility no other requirements are necessary, but this
requirement drastically restricts the choice of possible
models~[5], see also~[1,2].

The secondary fields~(1.1) exhibit anomalous transformation
properties with respect to internal and conformal symmetry groups
and satisfy the anomalous Ward identities~[1,2].

Each exactly solvable model is intrinsically related to a special
tensor field $Q_s^j(x)$ or $Q_s^T(x)$, $s = 1,2,\ldots$ of the
scale dimension $d+s$. This field is a superposition of secondary
fields
$$Q_s(x) = \sum_{s_1\ldots s_k} \alpha_{s_1\ldots s_k} P^{(s_1\ldots s_k)}_s(x), \quad
s_1\le s_2\le\ldots\le s_k, \quad s_1+s_2+\ldots+ s_k\le s-1,$$
where $\alpha_{s_1\ldots s_k}$ are the scale factors.  These
factors are defined by the condition that the anomalous terms in
Ward identities for the Green functions $\langle j_\mu
Q_s^j\varphi\ldots\rangle$ or $\langle T_{\mu\nu}
Q_s^T\varphi\ldots\rangle$ should vanish.  This is equivalent to a
set of requirements~[1,2]
\begin{equation}
\langle j_{\mu} Q_s^j\varphi\rangle = \langle j_{\mu}
Q_s^jP^j_{s'}\rangle = 0\quad\mbox{for all $s'$,}
\end{equation}
and analogously for the the energy-momentum tensor $T_{\mu\nu}$
and the field $Q_s^T$.

This being the case, the fields $Q_s^j$ or $Q_s^T$ have normal
transformation properties.  Each model is fixed by the requirement
that the states $Q_s^j(x)|0\rangle$ or $Q_s^T(x)|0\rangle$ vanish:
$$ Q_s^j(x) = 0 \mbox {\ or\ } Q_s^T(x)=0.$$
The dimension $d$ of the fundamental field $\varphi(x)$, as well
as the coefficients $\alpha_{s_1\ldots s_k}$, are calculated from
the equations~(1.4).

Let us briefly discuss this construction for $D=2$.  Notice that
for $D\ge3$ the conformal group is
$\frac{1}{2}(D+1)(D+2)$-parametric so we do not employ any analogs
of the Virasoro algebra.  Nevertheless, if one formally sets $D=2$
in the above construction, one may obtain~[1,2] the well known~[6]
family of two dimensional conformal models.  Traditionally, these
models are derived using the different construction~[6,7] based on
the Virasoro algebra.  However, as described in~[1], the exact
solution of such two dimensional models may be obtained without
the Virasoro algebra (i.e., using only the 6-parametric conformal
group) in the framework of the approach proposed herein.  In
particular, in $D=2$ the fields $P_s^T$ introduced above coincide
with the covariant superpositions of the secondary fields
$a_{-k_1}\ldots a_{-k_m}\varphi(x)$, where $a_{-k}, k>0$  are the
generators of the Virasoro algebra.  In $D=2$ the states
$Q_s(s)|0\rangle$ coincide with the null vectors of the two
dimensional conformal algebra, and the requirements~(1.4) result
in the $D$-dimensional analog of the Katz formula~[8].

In the general case, the fields~(1.1) for $D\ge 3$ may exist only
if the operator product expansions $j_\mu j_\nu$ and
$T_{\mu\nu}T_{\rho\sigma}$ include the anomalous terms~[1,2]
\begin{equation}
j_\mu(x) j_\nu(0) = [R_j] + \ldots, \quad
T_{\mu\nu}(x)T_{\rho\sigma}(0) = [R_T] + \ldots,
\end{equation}
where $R_j$ and $R_T$ are the scalar fields of dimensions
$$d_{R_j}=d_{R_T}=D-2.$$
In $D=2$, the scalar fields $R_j$ and $R_T$ are reduced to
constants which coincide with the central charges of the two
dimensional theory.

One of the most important goals of the approach proposed in~[1,2]
for $D\ge3$ is to establish the equations for the fields $R_j$ and
$R_T$, to find the conditions when the fields~(1.1) do exist, and
are well defined, and to calculate their Green functions. These
problems are, in principle, resolved in this article.

Below we demonstrate that the fields~(1.1) exist and have
well-defined Green functions only for selected values of
normalization constants of the Green functions
$\langle\varphi\varphi^+R_j\rangle$ and $\langle\varphi\varphi
R_T\rangle$.

For simplicity we calculate the constant for the function
$\langle\varphi\varphi^+R_j\rangle$ in the non-abelian
case\footnote{The abelian case is exceptional; it demands setting
$R_j=0$ and taking into account the $C$-number contribution into
the operator product expansion $j_\mu(x)j_\nu(0)$, see~[1,2].}.
The internal symmetry group generators $t^a$ are chosen to be
anti-hermitian
\begin{eqnarray}
&& (t^a)^+ = -t^a,\ [t^a,t^b]=f^{abc}t^c,\ t^at^a=-C_g,\ f^{abc}t^bt^c=C_vt^a\nonumber \\
&& (t^a\varphi)_\alpha = (t^a)^\beta_\alpha\varphi_\beta,\
   (t^a\varphi^+)^\alpha = -(\varphi^+)^\beta(t^a)^\alpha_\beta. \nonumber
\end{eqnarray}
The field $\varphi(x)=\varphi_{\alpha}(x)$ will be considered as a
scalar field.

 Analogous calculations for the energy-momentum
tensor are described only qualitatively.

\section{Requirements for Existence of Secondary Fields}
Consider the Green functions $\langle
P_s^j\varphi^+j^{a}_\mu\rangle$. They are calculated from the
equation~[1,2]
\begin{eqnarray}
&&\langle
P_s^j\varphi^+j^{a}_\mu\rangle \equiv\langle P^j_{\mu_1\ldots\mu_s}(x_1)\varphi^+(x_2)j^a_\mu(x_3)\rangle \nonumber \\
&&\qquad\quad{}= -\res_{l=d+s}\int dy_1\,dy_2\,
B^l_{\mu_1\ldots\mu_s}(x_1y_1y_2) t^b \partial^{y_2}_\nu \langle
j_\nu^b(y_2)\varphi(y_1)\varphi^+(x_2)j_\mu^a(x_3)\rangle,
\end{eqnarray}
where
\begin{eqnarray}
&& B^l_{\mu_1\ldots\mu_s}(x_1x_2x_3) =
\lambda^{x_1}_{\mu_1\ldots\mu_s}(x_2x_3)
\left(\frac{1}{2}x_{12}^2\right)^{-\frac{l-d-s+D}{2}}
\left(\frac{1}{2}x_{13}^2\right)^{-\frac{l+d-s-D}{2}}
\left(\frac{1}{2}x_{23}^2\right)^{\frac{l+d-s-D}{2}} \nonumber \\
&& \lambda^{x_1}_{\mu_1\ldots\mu_s}(x_2x_3) =
\lambda_{\mu_1}^{x_1}(x_2x_3)\ldots \lambda_{\mu_s}^{x_1}(x_2x_3) - {\rm traces,} \nonumber\\
&&\lambda_\mu^{x_3}(x_1x_2) = \frac{(x_{13})_\mu}{x_{13}^2} -
\frac{(x_{23})_\mu}{x_{23}^2}, \qquad (x_{ij})_{\mu} =
(x_{i})_{\mu} - (x_{j})_{\mu} .
\end{eqnarray}
To calculate the r.h.s.\ of Eq.(2.1) one should use the anomalous
Ward identity
\begin{eqnarray}
&&\partial^x_\nu \langle j_\nu^b(x)
j_\mu^a(y)\varphi(x_1)\varphi^+(x_2)\rangle =
-\sum_{i=1}^2\delta(x-x_i)t_{x{_i}}^b\langle j_\mu^a(y)\varphi(x_1)\varphi^+(x_2)\rangle \nonumber \\
&&\qquad{}+f^{bac}\delta(x-y) \langle
j_\mu^c(y)\varphi(x_1)\varphi^+(x_2)\rangle +
\partial_\mu^x\delta(x-y) \langle
R^{ab}(y)\varphi(x_1)\varphi^+(x_2)\rangle,
\end{eqnarray}
where the notation $t_{x{_i}}^a$ means that the matrix $t^a$ act
on the field $\varphi(x_{i})$ in accordance with
\begin{eqnarray}
&& (t^a\varphi)_\alpha = (t^a)^\beta_\alpha\varphi_\beta,\
   (t^a\varphi^+)^\alpha = -(\varphi^+)^\beta(t^a)^\alpha_\beta. \nonumber
\end{eqnarray}
 As the result, the expression under the ``res'' sign takes
the form
\begin{eqnarray}
&&\left(C_g-\frac{1}{2}C_V\right)\int dx_5\,
B^l_{\mu_1\ldots\mu_s}(x_1x_5x_2)
\langle\varphi(x_5)\varphi^+(x_2)j_\mu^a(x_3)\rangle\nonumber\\
&&\qquad{}+\frac{1}{2}C_V \int
dx_5\,B^l_{\mu_1\ldots\mu_s}(x_1x_5x_3)
\langle\varphi(x_5)\varphi^+(x_2)j_\mu^a(x_3)\rangle\nonumber\\
&&\qquad\qquad{}+t^b \int dx_5\,
\left[\partial_\mu^{x_3}B^l_{\mu_1\ldots\mu_s}(x_1x_5x_3)\right]
\langle\varphi(x_5)\varphi^+(x_2)R^{ab}(x_3)\rangle.
\end{eqnarray}
Conformal expressions for the functions
$\langle\varphi\varphi^+j_\mu^a\rangle $ and
$\langle\varphi\varphi^+R^{ab}\rangle $ are (see ~[1,2] ):
\begin{eqnarray}
&&\langle\varphi(x_1)\varphi^+(x_2)j_\mu^a(x_3)\rangle
=g_j t^a\lambda^{x_3}_\mu(x_1x_2)\Delta_j(x_1x_2x_3),\nonumber\\
&& t^b \langle\varphi(x_1)\varphi^+(x_2)R^{ab}(x_3)\rangle =g_R
t^a\Delta_j(x_1x_2x_3),
\end{eqnarray}
where $g_j$ and $g_R$ are the normalization constants, and
$$\Delta_j(x_1x_2x_3) =
\left(\frac{1}{2}x_{12}^2\right)^{-\frac{2d-D+2}{2}}
\left(\frac{1}{2}x_{13}^2\right)^{-\frac{D-2}{2}}
\left(\frac{1}{2}x_{23}^2\right)^{-\frac{D-2}{2}}.$$ The constant
$g_j$ is calculated from the Ward identity for the function
$\langle\varphi\varphi^+j_\mu^a\rangle$; it equals to
$g_j=(2\pi)^{-D/2}\Gamma(D/2)$.

Eventually, for the Green function $\langle
P^{j}_s\varphi^+j^{a}_\mu\rangle$ in the l.h.s. of Eq.~(2.1) we
must derive a conformally invariant expression which depends on
the normalization constants $g_j,g_R$. A general expression for
the Green function $\langle P^{j}_s\varphi^+j^{a}_\mu\rangle$
is~[1]
\begin{equation}
\langle P_s\varphi^+j_\mu^a\rangle = \lim_{l=d+s}
\left[A_1(l,s)C^l_{1\mu,\mu_1\ldots\mu_s}(x_1x_2x_3)+
A_2(l,s)C^l_{2\mu,\mu_1\ldots\mu_s}(x_1x_2x_3)\right],
\end{equation}
where
\begin{eqnarray}
&&\kern-10pt C^l_{1\mu,\mu_1\ldots\mu_s}(x_1x_2x_3) =
\lambda_\mu^{x_3}(x_1x_2)\lambda^{x_1}_{\mu_1\ldots\mu_s}(x_3x_2)
\Delta_j^l(x_1x_2x_3),\nonumber\\
&&\kern-10pt C^l_{2\mu,\mu_1\ldots\mu_s}(x_1x_2x_3) =
\frac{1}{x_{13}^2}\left[ \sum^s_{k=1}g_{\mu\mu_k}(x_{13})
\lambda^{x_1}_{\mu_1\ldots\hat{\mu}_k\ldots\mu_s}(x_3x_2)-{\rm
traces\ }\mu_1\ldots\mu_s
\right]\Delta_j^l(x_1x_2x_3),\nonumber\\
&&\kern-10pt g_{\mu\nu}=\delta_{\mu\nu}-2\frac{x_\mu
x_\nu}{x^2},\quad \hat{\mu}_k \quad \mbox{means the omission
of the index,} \nonumber\\
&&\kern-10pt \Delta_j^l(x_1x_2x_3) =
\left(\frac{1}{2}x_{12}^2\right)^{-\frac{l+d-s-D+2}{2}}
\left(\frac{1}{2}x_{13}^2\right)^{-\frac{l-d-s+D-2}{2}}
\left(\frac{1}{2}x_{23}^2\right)^{-\frac{-l+d+s+D-2}{2}},\nonumber
\end{eqnarray}
and the coefficients $A_i(l,s), i=1,2,$ should be calculated from
Eq.~(2.4) and depend on the constants $g_R,g_j$.

It is essential that both terms in Eq.~(2.6) are poorly defined in
the limit $l=d+2$ because of the factor
$\left(\frac{1}{2}x_{13}^2\right)^{-\frac{l-d-s+D}{2}}$ which
appears in the contractions over $\mu,\mu_k$, $k=1,\ldots,s$. As
shown in~[1], for $l=d+s$ their residues have the form
\begin{eqnarray}
&&(D-2) \res_{l=d+s}C^l_{1\mu,\mu_1,\ldots\mu_s}(x_1x_2x_3) = -
\res_{l=d+s}C^l_{2\mu,\mu_1,\ldots\mu_s}(x_1x_2x_3)\nonumber\\
&&\qquad\quad{}\sim \sum_{k=1}^s
\biggl[(-2)^{-k}\frac{1}{k!}\delta_{\mu\mu_1}
\partial_{\mu_2}^{x_3}\ldots\partial_{\mu_k}^{x_3}\delta(x_{13})
\frac{(x_{12})_{\mu_{k+1}}}{x^2_{12}} \ldots
\frac{(x_{12})_{\mu_s}}{x^2_{12}}\nonumber\\
&&\qquad\qquad\quad{}+\{\ldots\}- {\rm traces\
}\mu_1\ldots\mu_s\biggr] (x_{12}^2)^{-d},
\end{eqnarray}
where $\{\ldots\}$ are the symmetrizing permutations of the
indices $\mu_1\ldots\mu_s$.  Hence the non-integrable
singularities in Eq.~(2.6) cancel if the following condition is
satisfied:
\begin{equation}
\lim_{l=d+s}{A_1(l,s)\over A_2(l,s)}=D-2.
\end{equation}
In this case, the Green function $\langle P_s\varphi^+
j_\mu^a\rangle$ takes the form~[1]
\def\lrpar{\mathop{\partial}\limits^\leftrightarrow{}}
\def\lpar{\mathop{\partial}\limits^\leftarrow{}}
\def\rpar{\mathop{\partial}\limits^\rightarrow{}}

\begin{equation}
\langle P_s(x_1)\varphi^+(x_2)j^a_\mu(x_3)\rangle \sim
t^a(x_{23}^2)^{-\frac{D-2}{2}}\lrpar_{\!\!\mu}^{\!x_3}\left[
(x_{13}^2)^{-\frac{D-2}{2}}\lambda^{x_1}_{\mu_1\ldots\mu_s}(x_3x_2)
\right](x_{12}^2)^{-\frac{2d-D+2}{2}}+{\rm [q.t.],}
\end{equation}
where $\lrpar=\rpar-\lpar$ and the quasi-local
terms [q.t.] are given by the expression (2.7). In what follows,
the condition~(2.8) will be considered as the requirement for
existence of the fields $P_s$.

Let us substitute (2.5) into (2.4). The first term was calculated
in~[1] and was shown to satisfy the criterion (2.8). The sum of
the second and third terms is calculated using the integral
relations listed in~[1]. The result is
\begin{eqnarray}
&& t^a\frac{1}{2}g_j(2\pi)^{D/2}
\frac{\Gamma\left(\frac{d+s-l}{2}\right)}{\Gamma\left(\frac{l-d+s+D}{2}\right)}
(-1)^{s+1}
\frac{\Gamma(D-d)\Gamma\left(\frac{l+d+s-D}{2}\right)}{\Gamma(d-D/2+1)
\Gamma\left(\frac{2D-l-d+s}{2}\right)}\nonumber\\
&&\qquad{}\times\Biggl\{-\Biggl[\frac{1}{2}C_V(l+d+s-D)+
\frac{g_R}{g_j}\frac{1}{2}(d+s-l)\nonumber\\
&&\qquad\qquad{} \times\frac{(l+d+s-D)(l+d-s-D)}{2(D-d-1)}\Biggr]
C^l_{1\mu,\mu_1,\ldots\mu_s}(x_1x_2x_3)\nonumber\\
&&{}+\Biggl[{1\over2}C_V-\frac{g_R}{g_j}\Biggl[(l+d-s-D)+
\frac{(l-d+s+D-2)(2D-l-d+s-2)}{2(D-d-1)}\Biggr]\Biggr]\nonumber\\
&&{}\times
C^l_{2\mu,\mu_1,\ldots\mu_s}(x_1x_2x_3)\Biggr\}.\nonumber
\end{eqnarray}
Now let us calculate a residue of this expressions at the points
$l=d+s$. It is readily seen that the result satisfies the
criterion~(2.8) for all $d$ and $s$ if we set
\begin{equation}
g_R=\frac{C_V}{2(D-2)}g_j=\frac{C_V}{4}(2\pi)^{-D/2}\Gamma\left(
\frac{D-2}{2}\right).
\end{equation}
Thus the complete family of the secondary fields $P_s$ which
appears in the operator product expansion $j_\mu^{a}\varphi$ has
well defined Green functions~(2.9), provided that the operator
product expansion $j_\mu^a(x)j_\nu^b(0)$ includes the field
$R_j^{ab}$ and the normalization of the Green function $\langle
R_j^{ab}\varphi\varphi^+\rangle$ is given by Eq.~(2.10).

\section{A Family of Conformal Fields with Canonical \\ Dimensions}
Consider the operator product expansion $j_\mu^a R^{bc}$. It
involves the tensor fields $R_s$ which are analogous to the fields
$P_s$. Apparently, the dimensions of the fields $R_s$ are obtained
by the substitution $d\to D-2$ in~(1.2). We examine the fields
which transform as vectors with respect to their inner index. So
we have the family of the secondary fields
\begin{equation}
R_s=R^a_{\mu_1\ldots\mu_s}(x),\quad d_s=D-2+s
\end{equation}
which arises in the operator expansion
\begin{equation}
j_\mu^a(x)R^{bc}(0) = [R^{ab}]+\sum_{s\ge1}[R_s^q].
\end{equation}
The current $j_\mu^a$ belongs to this family:
$j_\mu^a(x)=R_s^a|_{s=1}$.

The Green functions of the fields $R_s^{a}$ are obtained from the
expressions~(2.6) after the substitution $d\to D-2$, while the
existence criterion of these functions comes up when the same
substitution is performed in Eq.~(2.8). To analyze the Green
functions it proves convenient to extract the traceless part
$\hat{R}^{ab}$ of the field $R^{ab}$
$$R^{ab} = \hat{R}^{ab}+{1\over N}\delta^{ab}R,\quad R= R^{aa},$$
and apply the Ward identities
\begin{eqnarray}
&&\partial_\nu^{x_4} \langle
j_\nu^m(x_4)j_\mu^n(x_3)\hat{R}^{ab}(x_1)\hat{R}^{cd}(x_2)\rangle=
\delta(x_{34}) f^{mnk}
\langle j_\mu^k(x_3) \hat{R}^{ab}(x_1)\hat{R}^{cd}(x_2) \rangle \nonumber\\
&&\qquad\quad{}+ \delta(x_{14}) f^{mak}
\langle j_\mu^n(x_3)\hat{R}^{kb}(x_1)\hat{R}^{cd}(x_2)\rangle + (a\leftrightarrow b)\nonumber \\
&&\qquad\quad{}+ \delta(x_{24}) f^{mck}
\langle j_\mu^n(x_3)\hat{R}^{ab}(x_1)\hat{R}^{kd}(x_2)\rangle + (c\leftrightarrow d)\nonumber \\
&&\kern-20pt{}+\partial_\nu^{x_4}\delta(x_{34}) \langle
\hat{R}^{mn}(x_3)R^{ab}(x_1)\hat{R}^{cd}(x_2) \rangle + {1\over N}
\delta^{mn}
\partial_\nu^{x_4}\delta(x_{34})
\langle R(x_3)R^{ab}(x_1)\hat{R}^{cd}(x_2)\rangle \\
&&
\partial_\nu^{x_4}
\langle j_\nu^m(x_4)j_\mu^n(x_3)R^{ab}(x_1) R(x_2)\rangle=
\partial_\nu^{x_4}\delta(x_{34})
\langle R^{mn}(x_3)R^{ab}(x_1)R(x_2)\rangle.
\end{eqnarray}
Let us find out the Green function $\langle
R_s^a\hat{R}^{cd}j_\mu^n\rangle$. It is determined by the equation
analogous to Eq.~(2.1):
\begin{eqnarray}
&& \langle R^q_{\mu_1\ldots\mu_s}(x_1)\hat{R}^{cd}(x_2)j_\mu^n(x_3)\rangle\nonumber \\
&&{}=-\res_{l=D-2+s}\int dx_4\,dx_5\, B^{q\
abm}_{\mu_1\ldots\mu_s}(x_1x_5x_4)
\partial_\nu^{x_4}
\langle
j_\nu^m(x_4)j_\mu^n(x_3)R^{ab}(x_5)\hat{R}^{cd}(x_2)\rangle,
\end{eqnarray}
where $B^{q\ abm}_{\mu_1\ldots\mu_s}(x_1x_5x_4) =
(\delta^{am}\delta^{bq}+\delta^{aq}
\delta^{bm})\hat{B}^l_{\mu_1\ldots\mu_s}(x_1x_5x_4)$, and
$\hat{B}^l_{\mu_1\ldots\mu_s}$ is obtained from the
expressions~(2.2) after the substitution $d\to D-2$.

To simplify the calculation, consider the case of the group
$SU(2)$. We put $f^{abc}=\epsilon^{abc}$, $\delta^{aa}=N=3$, where
$\epsilon^{abc}$ is the totally antisymmetric tensor. The Green
functions in the r.h.s.\ of the Ward identity~(3.3) read
\begin{eqnarray}
&& \langle j_\mu^n(x_3)\hat{R}^{ab}(x_1)\hat{R}^{cd}(x_2)\rangle =
h_j\left[f^{nac}\delta^{bd}+f^{nbc}\delta^{ad}+(c\leftrightarrow
d) \right]
\lambda_\mu^{x_3}(x_1x_2)\Delta_R(x_1x_2x_3), \nonumber \\
&& \langle \hat{R}^{ab}(x_1)\hat{R}^{cd}(x_2) R(x_3)\rangle =
h_1\left[\delta^{ac}\delta^{bd}+\delta^{ad}\delta^{bc}-{2\over3}\delta^{ab}\delta^{cd}\right]
\Delta_R(x_1x_2x_3), \nonumber \\
&& \langle
\hat{R}^{ab}(x_1)\hat{R}^{cd}(x_2)\hat{R}^{mn}(x_3)\rangle =
h_2\bigl[\delta^{ac}\delta^{bm}\delta^{dn}+(a\leftrightarrow b)\nonumber \\
&& \qquad\qquad{}+ (c\leftrightarrow d)+(m\leftrightarrow n) -{\rm
traces}\biggr]\Delta_R(x_1x_2x_3), \nonumber
\end{eqnarray}
where $h_1,h_2,h_3$ are the constants, $\Delta_R(x_1x_2x_3) =
({1\over8}x_{12}^2x_{13}^2x_{23}^2)^{-\frac{D-2}{2}}$. Substitute
these equations, as well as the identity~(3.3), into Eq.~(3.5).
The expression of the residue part will be analogous to that
in~(2.4). Taking the integrals with the help of relations provided
in~[1] we find the following result: the r.h.s.\ of Eq.~(3.5)
satisfies the criterion of existence (see~(2.6),(2.8) for $d\to
D-2$) if
\begin{equation}
(D-2)\left({4\over3}h_1+{14\over3}h_2\right)+3h_j = 0.
\end{equation}
When this requirement holds, the Green functions $\langle
R_s\hat{R}j_\mu\rangle$ have the form\footnote{For even $D$ these
expressions demand more precise definition. This problem will be
addressed in the other article.}:
\begin{eqnarray}
&& \langle
R^q_{\mu_1\ldots\mu_s}(x_1)\hat{R}^{ab}(x_2)j_\mu^n(x_3) \rangle
\sim(\delta^{aq}\delta^{bn}+\delta^{an}\delta^{bq}-{1\over N}\delta^{ab}\delta^{qn})\nonumber\\
&&{}\times(x^2_{23})^{-\frac{D-2}{2}}
\lrpar_{\!\!\mu}^{\!x_3}\left[
(x_{13}^2)^{-\frac{D-2}{2}}\lambda^{x_1}_{\mu_1\ldots\mu_s}(x_3x_2)\right]
(x_{12}^2)^{-\frac{D-2}{2}}+\mbox{\it quasilocal terms.}
\end{eqnarray}
Using the Ward identity~(3.4) one can show that the Green
functions $\langle R^q_{\mu_1\ldots\mu_s}Rj^a_\mu\rangle$ where
$R=R^{aa}$ contain only quasilocal terms which are given by the
expression~(2.7) after the substitution $d\to D-2$.

Consider the higher Green functions of the fields $R_s$. They are
determined from the equation
\begin{eqnarray}
&&\langle R^a_s(x)\varphi(x_1)\ldots\varphi^+(x_{2n})\rangle =
-(\delta^{ac}\delta^{bd}+
\delta^{ad}\delta^{bc}) \nonumber \\
&&{}\times\res_{l=D-2+s}\int dy_1\,dy_2\,
\hat{B}^l_{\mu_1\ldots\mu_s}(xy_1y_2)\partial^{y_2}_\nu \langle
j_\nu^b(y_2)R^{cd}(y_1)\varphi(x_1)\ldots\varphi^+(x_{2n})\rangle.
\end{eqnarray}
The method used to calculate the r.h.s.\ is described in detail
in~[1], see also~[2].

Consider this equation for $s=1$. As mentioned above, the field
$R_s^a$ for $s=1$ coincides with the current $j_\mu^a$.  Using the
methods developed in~[1] one can show that for $s=1$ Eq.~(3.8)
leads to the following relation (taking into account that
$R_s^a|_{s=1}=j_\mu$):
\begin{equation}
\langle j_\mu^a(x)\varphi(x_1)\ldots\varphi^+(x_{2n})\rangle =
\gamma \sum^{2n}_{k=1} n\frac{(x-x_k)_\mu}{(x-x_k)^2}t_{x_{k}}^b
\langle R^{ab}(x)\varphi(x_1)\ldots\varphi^+(x_{2n})\rangle.
\end{equation}
The constant $\gamma$ is calculated from the same equation for
$n=1$. Taking into account~(2.10), we have:
$\gamma=g_j/g_R=\frac{C_V}{2(D-2)}$. The relation~(2.9) and the
Ward identity for the Green functions $\langle
j_\mu\varphi\ldots\rangle$ lead to the following equation
\begin{eqnarray}
\gamma\partial_\mu^x\sum_{k=1}^{2n}\frac{(x-x_k)_\mu}{(x-x_k)^2}t_{x_{k}}^b
\langle R^{ab}(x)\varphi(x_1)\ldots\varphi^+(x_{2n})\rangle =
-\sum_{k=1}^{2n}\delta(x-x_k) t_{x_{k}}^a \langle
\varphi(x_1)\ldots\varphi^+(x_{2n})\rangle
\nonumber\\
\end{eqnarray}

\section{Secondary Fields Generated by\hfil\break the Energy-Momentum Tensor}
The results analogous to those described above may be obtained for
the secondary fields generated by the energy-momentum tensor. Let
us consider them qualitatively.

The requirement for existence of the fields $P_s$ coincides with
the condition that the Green functions $\langle P_s^T\varphi
T_{\mu\nu}\rangle$ are well defined. For $D\ge3$, the general
conformally invariant expression of the functions $\langle P_s^T
\varphi T_{\mu\nu}\rangle$ has the form:
\begin{equation}
\langle P_s^T(x_1) \varphi(x_2) T_{\mu\nu}(x_3)\rangle =
\lim_{l=d+s} \left[\sum_{i=1}^3
A_i(l,s)C_{i,s}^l(x_1x_2x_3)\right],
\end{equation}
where $C_{i,s}^l(x_1x_2x_3)$ are the basic conformal
functions~[1,2] and $A_i$ are arbitrary coefficients.  All the
functions $C_{i,s}^l$ are poorly defined for $l=d+s$. However, the
limit of the sum in~(4.1) is a well defined function provided
that~[1,2]
\begin{equation}
\lim_{l=d+s}\frac{A_1(l,s)}{A_2(l,s)} = D(D-2),\quad
\lim_{l=d+s}\frac{A_2(l,s)}{A_3(l,s)} = D.
\end{equation}
The Green functions $\langle P_s^T \varphi T_{\mu\nu}\rangle$ must
be found from the equation analogous to Eq.~(2.1). Its r.h.s.\
includes the quantity $\partial_\mu\langle
T_{\mu\nu}T_{\rho\sigma}\varphi\varphi\rangle$ determined by an
anomalous Ward identity. This Ward identity was derived in~[1]
(see also~[2]). The calculations are carried out in the same
manner as in the case of the current. As the result, the
criterion~(4.2) leads to certain conditions which impose the
relations between the normalization of the Green functions
$\langle R_T\varphi\varphi \rangle$ (where $R_T$ is the anomalous
contribution in the expansion $TT=[R_T]+\ldots$) and of the other
free parameters~[1,2] entering the anomalous Ward identity for
$\langle T_{\mu\nu}T_{\rho\sigma}\varphi\varphi\rangle$.

The operator product expansion
$$T_{\mu\nu}(x)R_T(0)=\sum_s[R_s^T]$$
where $R_s^T=R_{\mu_1\ldots\mu_s}$ are the tensor fields of
canonical dimensions $D-2+s$, is handled analogously. The
condition of their existence is derived from Eqs.~(4.1) and~(4.2)
after the substitution $d\to D-2$. Identically to the case of the
current, this condition fixes up the normalization of the Green
function $\langle R_TR_TR_T\rangle$ which enters the anomalous
Ward identity for $\langle T_{\mu\nu}
T_{\rho\sigma}R_TR_T\rangle$. The field $R_T(x)$ is the
fundamental field of the family of fields $R_s^T$. The
energy-momentum tensor belongs to this family:
$T_{\mu\nu}(x)=R_s^T|_{s=2}$. Using this, one can express the
Green functions $\langle T_{\mu\nu}\varphi\ldots\varphi\rangle$ in
terms of the functions $\langle R\varphi\ldots\varphi\rangle$, and
to find the equation for the functions $\langle
R\varphi\ldots\varphi\rangle$, analogously to the case of
Eqs.~(3.9),(3.10). All necessary calculations will be presented
elsewhere.

\section*{Acknowledgments}

 The work was supported in part by
grants RFBR No. 05-02-17654, LSS No. 1578.2003-2 and INTAS No.
03-51-6346.


\begin{thebibliography}{8}
\bibitem{1} E.S.Fradkin, M.Ya.Palchik, {\it Conformal Quantum Field
Theory,\/} Kluwer Acad. Publ., 1996
\bibitem{2} E.S.Fradkin, M.Ya.Palchik, {\it Phys. Rep.\/} {\bf 300}
(1998) 1; {\it Int. Journ. of Mod. Phys.\/} {\bf 13} (1998) 4787;
{\bf 13} (1998) 4837
\bibitem{3} V.K.Dobrev, G.Mack, V.B.Petkova, S.G.Petkova, I.T.Todorov,
{\it Lecture Notes in Physics,\/} v.63 (Springer-Verlag, 1977).
\bibitem{4} A.U.Klimyk, A.M.Gavrilik, {\it Matrix elements and Klebsh-Gordon
Coefficients of Group Representations\/} (Naukova Dumka, Kiev,
1979)
\bibitem{5} A.A.Belavin, A.M.Polyakov, A.B.Zamolodchikov,
{\it Nucl. Phys.\/} {\bf B241} (1984) 333
\bibitem{6} V.G.Knizhnik, A.B.Zamolodchikov, {\it Nucl. Phys.\/}
{\bf B247} (1984) 333
\bibitem{7} E.S.Fradkin, M.Ya.Palchik, preprint IASSNS-Hep-97/122,
Princeton, 1997, hep-th/9712045
\bibitem{8} V.G.Kac, {\it Lecture Notes in Physics\/} {\bf 94} (1970) 120
\end{thebibliography}
\end{document}